\documentclass[namedreferences]{solarphysics}
\usepackage[hyperref,optionalrh,showbiblabels]{spr-sola-addons} 
\usepackage{graphicx}        
\usepackage{color}           
\usepackage{breakurl}        

\chardef\us=`\_

\begin{document}
\begin{article}
\begin{opening}
\title{Multi Photon Ionization in Gravity Induced Resonant Emission}
\author [corref, email={antonysoosaleon@yahoo.com}]{\inits{S. Antony}\fnm{Antony Soosaleon}}
\address {School of Pure \& Applied Physics, Mahatma Gandhi University, Kottayam, Kerala, India-686560}
\runningauthor{Antony Soosaleon}
\runningtitle{Example paper}
\begin{abstract} 
		\begin{figure}
		\centering
		\includegraphics[height=5.5cm]{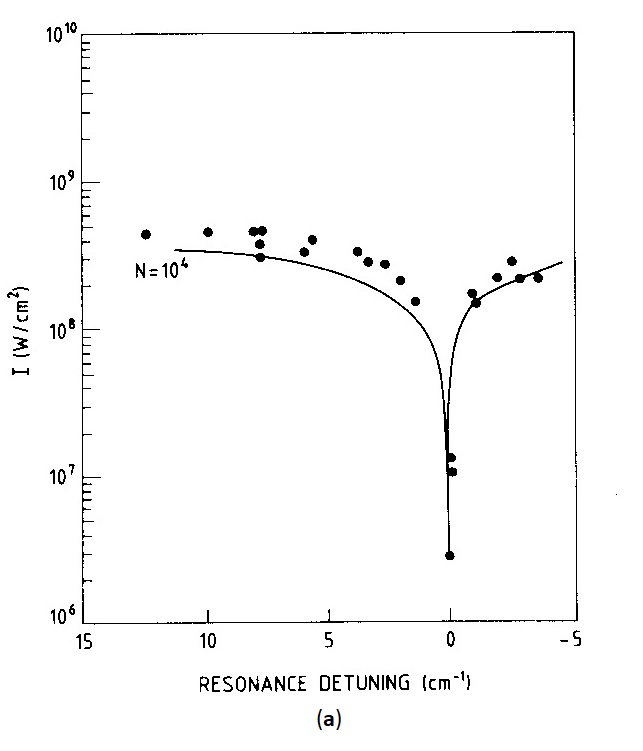}
		\includegraphics[height=5.5cm]{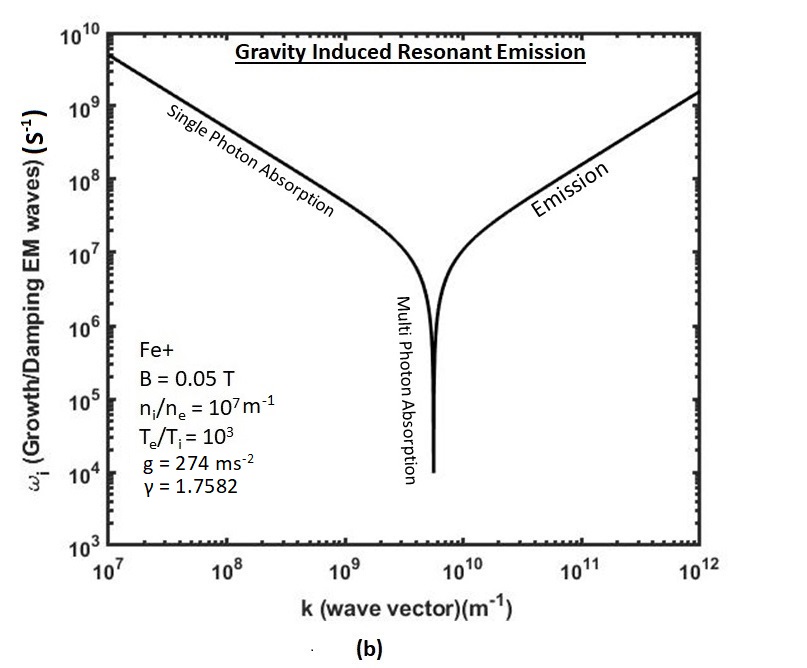}
		\caption{(a)(Courtesy : \citet{Mainfray1980}) Comparative plot of Experimental \citep{Morellec1976} and theoretical calculation (solid line) \citep{Crance1978} of 4 photon Ionization for Cs ; (b) Logarithemic plot of resonance condition at \emph{\textbf{gire}} for Fe+ ions in coronal plasma. Both the graph shows absorption/emission of \emph{\textbf{emws}} with respect to the resonance condition}.
	\end{figure}

	Gravity Induced Resonant Emission (\emph{\textbf{gire}}) has been proposed as the mechanism of coronal heating; here we are comparing this with Multi photon Ionization (\emph{\textbf{mpi}}) in Laser. We find  both are same, except electron damping as a prephase mechanism to initiate \emph{\textbf{mpi}} in \emph{\textbf{gire}}. A high intensity laser field used for \emph{\textbf{mpi}}, but in \emph{\textbf{gire}} this happens at resonance condition when the Larmour radii of the particles are same. The temporal compression of photons is achieved by grating devise, but in \emph{\textbf{gire}}, this is done by field deficit created due to the loss of electrons at resonance condition and this loss is compensated by photo electrons emitted during \emph{\textbf{mpi}}. Fig.1 (a) \& (b) are plots of \emph{\textbf{mpi}} in Laser and resonance condition of \emph{\textbf{gire}}, shows absolute resemblance except the wider range of frequency spectrum in \emph{\textbf{gire}}.  \\
	
 \end{abstract}
\end{opening}
\section{Introduction}
    \label{S-Introduction} 
    
     Damping of electromagnetic waves (\emph{\textbf{emws}}) by cyclotron resonance of particles in the Lower Hybrid Oscillations (\emph{\textbf{lhos}}) has been proposed as the mechanism of \emph{\textbf{gire}}.  \emph{\textbf{lhos}} is a coupled oscillations, between ions and electrons in the cyclotron motion bound by Couloumb force. Gravitational drift of ions induces an electric field, causes this oscillations in a collisionless plasma where the magnetic field is controlling the motion of particles. Particles are mutual accelerated by this oscillations, where the necessary energy is absorbed from \emph{\textbf{emws}} by creating a field deficit between the particles. We have briefly discussed about the role of Landau damping over \emph{\textbf{gire}} in the previous papers \citep{Antony2017a, Antony2017b}; here we are going to analyze it on the basis of \emph{\textbf{mpi}} and through out this study, we are using the same growth rate and resonant frequency expression from the above 2 papers. 
This analysis is done in the following way:
 \begin{enumerate}
 		\item Initially we are briefing about the required conditions for \emph{\textbf{mpi}} in Laser
 		\item Describing how these conditions are satisfied in \emph{\textbf{gire}}
 		\item Discussing the effect of various parameters over \emph{\textbf{mpi}} in \emph{\textbf{gire}}
 	 \end{enumerate}
  
\section{Multi Photon Ionization}

\subsection{Photo Ionization} Ionization of atoms is possible with photons of energy greater than required, but \citet{Goeppert-Meyer1932} theoretically predicted that it is possible with photons of lesser energy. In 1961, Kaiser experimentally studied double photon ionization, followed by \citet{Abella1962} and \citet{Hall1965}. High power laser radiation was necessary for such ionization process and this led to the development of ultra fast Lasers. Multiple ionization was, first observed by \citet{Voronov1965} and \citet{Agostini1987} at low pressure of the order $10^{16} m^{-3}$.  Absorbing a large number of photons more than the required threshold value, results in multiple ionization. Such absorbtion is not possible for a free ion, since it violates momentum conservation. But this is can be made possible when the ion is at bound with some potential of other ions/electrons. In such condition, the momentum change due to the absorption of photons will be compensated with reverse momentum of bound particles. This condition is created in Femto/atto second lasers, the power of a laser is made intense more than the inter atomic field $10^{19} Wm^{-2}$, so that it can easily ionizes atoms at initial stage, in such a way that the emitted electrons stand in potential of the parant ion to give a recoiling momentum for further absorption of photons. Here the condition is that the power of laser must be sufficiently high, so that absorption of photons is possible before emitted electrons leave the potential of parant ion. This means absorption time is reduced while increasing the power of laser, which is named as temporal compression of photons. 

\subsection{Temporal Compression} There are more that the temporal Compression is a necessity for higher ionization; since ionized states are highly uncertain and the time periods are in the range 100-150 attoseconds. Ionizing atoms in such a short interval of time is not possible in ordinary situation, unless we reduce the absorption time of photon. Classically this cannot be less than the interaction time between electron and photon, whch are in femto seconds. These reasons makes clear that to achieve emph{\textbf{mpi}}, the absorption time must be reduced and this is done by chirpped pulse amplification in laser \citep{Donna1985}. Here spatial distribution of photons are reduced to produce high intensity pulse which gives the effect of temporal compression during absorption. This is achieved by grating element, which gives a definite phase shift to various photons and make them to converge at a single point to produce a very high intensity pulse. With a high intensity pulse greater than the interatomic field, a large number of photons can be absorbed by ions at the time of single photon absorption. This is a nonlinear process both temporally and spatially, but made possible in a high field condition.  Multiple ionization was first observed by \citep{L'Huillier1982} and such ionization results to the emission of UV and X-rays, the intensity of pulsed laser used for multiple ionization was about $10^{22} Wm^{-2}$, which is much larger than the inter-atomic field $10^{19} Wm^{-2}$. With these brief description, we shall list the required conditions for emph{\textbf{mpi}}.

\textbf{Four Conditions For Multi Photon Ionization}
\begin{enumerate}
	\item 	A high vacuum of the order $10^{16}m^{-3}$ is required to make collisionless 
	\item 	Particles must be in bound state with some potential to conserve the momentum 
	\item 	Intensity of Laser field must be more than the interatomic field ($10^{19} Wm^{-2}$)
	\item   Temporal compression of photons is required to reduce absorption time
\end{enumerate} 

All these 4 conditions are satisfied at resonance in \emph{\textbf{gire}}, we shall analyze in the following section.

\begin{figure}
	\includegraphics[height=5.3cm]{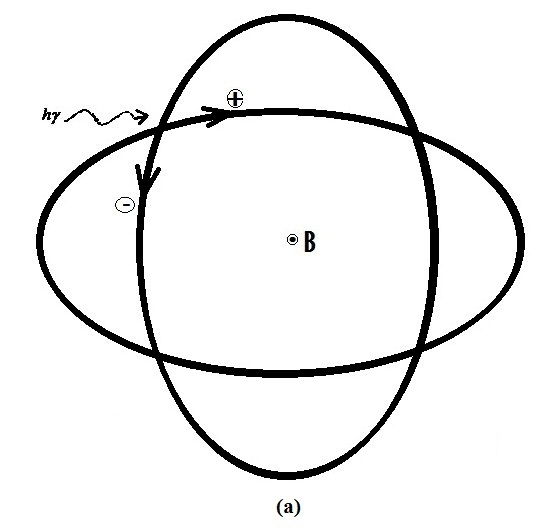}
	\includegraphics[height=5.5cm]{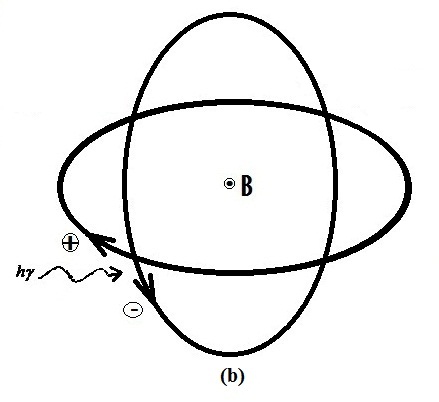}
	\caption{\textbf{Field Deficit Conditions for Elelctron\ Ion Landau Damping}: Fig. (a) \& (b) shows the sketch of field deficit condition for Ion \& Electron Landau damping. (a) shows the positions of coupled particles where incoming photons are absorbed by ion; here the field deficit is caused by motion of electrons, so that the rate of ion damping is at electron cyclotron frequency. (b) shows the positions particles where electron absorbs photon and here the field deficit is caused by motion of ions, so that the electron damping is at the rate of ion cylcotron frequency. Due to this, electron damping is a low energy (frequncy) prephase one but ion damping is a high energy and post phase one occurs at resonant condition (intense field)}.
\end{figure} 

\section{Conditions of \emph{mpi} in  \textbf{gire}}

The first 2 conditions need not much explanations, they are explicit in the theory of \emph{\textbf{gire}}, therefore we only marginally mentions these two.

\textbf{Condition 1: Corona is more vacuum \& Collisionless} : These are pre-assumed conditions in developing the theory of \emph{\textbf{gire}}, where the plasma was considerd as collisionless. The density used in the \emph{\textbf{gire}} plot was  $10^{15}m^{-3}$, which is more than the vacuum we have created in the Laboratory for Laser ionization. 

\textbf{Condition 2: Ions  \& Electrons are bound by Couloumb Force in \emph{lhos}}: Particles must be bound state with some potential to conserve the momentum during photon absorption. In laser ionization, this condition is created with photo electrons emited from initial ionization. But in \emph{\textbf{gire}}, this condition is already exisiting in \emph{\textbf{lhos}}, here the particles are bound by Couloumb force and there is more flexibility in reserving momentum since they are further controlled by magentic force.

\textbf{Condition 3: High Intensity Field; Particles are at intense electric field in resonance condition and this happens when Larmour radii of electrons and ions are same.}

\textbf{Condition 4: Temporal Compression of Photons; this is created by field deficit due to the loss of electrons at resonance condition} 

For understanding the last 2 conditions, we need a thorough discussion about the role of electron and ion Landau damping in \emph{\textbf{gire}}, after that we will be analysing these two in the sections 3.4 \& 3.5

\subsection{Role of Electron \& ion Landau Damping in \emph{gire}} 
Landau damping is a categorical phrase used in plasma physics for photon absorption process, named after Landau, who theoretically predicted this. We have analyzed this mechanism of energy transfer on the basis of field deficit in the  \emph{\textbf{lhos}}\citep{Antony2017b} by Couloumb force.  The photon absorption is a necessity of the particles which are in field deficit condition; this happens when they mutually cross over in the \emph{\textbf{lhos}}. There are 4 possible positions in the cyclotron orbits, where field deficit situation can arise, but we can eliminate 2 positions in considering the direction of photons.  In the rest 2, one is prephase, relative to the maximum amplitude (Electric field of incoming wave ) where the electron damping occurs (see the fig.2(b)), and the other is postphase, where the ion damping happens (fig.2(a)). Distinguishing these 2 are necessary to understand the coronal heating problem and both are equally important. We will be differentiating them on the basis of their fundemental properties and also compare our analysis with \emph{\textbf{gire}}.

\begin{figure}
	\centering
	\includegraphics[height=8cm]{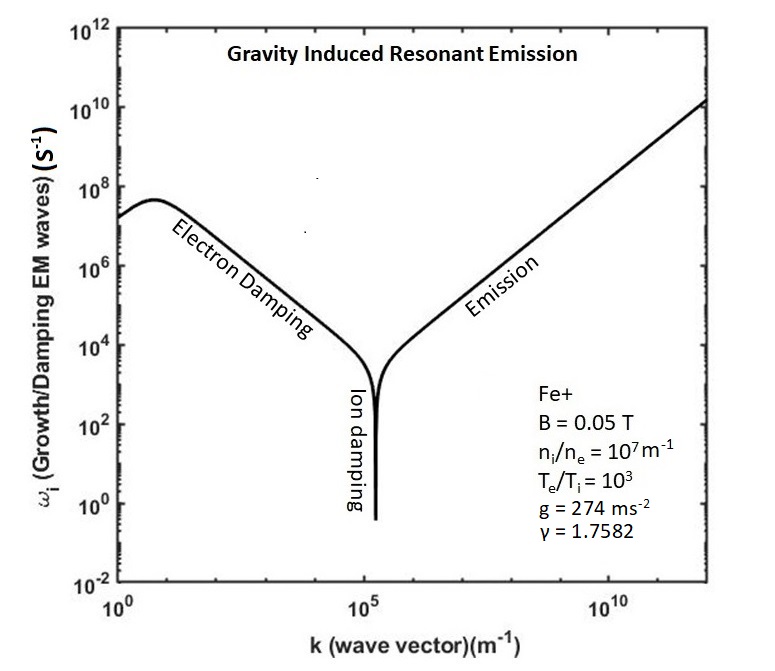}
	\caption{\textbf{Damping of \emph{emws} in two componant plasma of Fe+ for coronal condition}: This is logarithemic plot of  $\omega_i$ (Damping/growth of \emph{\textbf{emws}}) with wave vector (frequency). Already this graph has been analysed in the paper \citep{Antony2017a} but here we doing it again on the basis of photon absorption. The first part shows growth of \emph{\textbf{emws}} (+ve slope), this is due to damping of accoustic waves and the next part shows \emph{\textbf{emws}} are losing energy (-ve slope). This is a low frequency region and prephase damping due to the absorption of photon by electrons. This is a single photon absorption region since the frequency increases with damping. Next is the last part of damping before emission, this is due to high field resonance happens when Larmour radii are same. Here the wave vector is not changing with damping, i.e., a large number photons of same frequency is absorbed, a huge amount of energy and last part of damping before the emission must be ion Landau damping, which is similar to Multi photon Ionization in Laser}.
\end{figure}

\begin{table}
	\caption{Comparison of Electron \& Ion Landau Damping}
	\label{T-simple}
	\begin{tabular} {lcc}
		\hline 
		Electron Landau damping  & Ion Landau Damping \\
		\hline \hline
		It is a prephase damping & Postphase damping \\
		It is possble with low energy waves &  Only by high energy waves \\
		Damping occurs at a rate of $\Omega_i$ & At a rate of $\Omega_e$\\
		Acceleration is limited upto 500 KeV &  unlimitted \\
		Electrons are lost out during damping & Ionized to high charge state \\
		Electron density decreases & No change in density \\
		Happens at weak field condition & Only at resonance condition\\
		Single photon absorption & Multi photon absorption \\
		\hline
	\end{tabular}
\end{table}

\subsection{Electron Landau Damping}
This is a prephase damping and the positions of ions and electrons are depicted in fig.2(b). Here the motion of ions creates the field deficit to electrons and therefore electron damping is at the rate of ion cyclotron frequency ($\Omega_i$). In the same way, ion damping is a post phase one and the positions are shown in the fig.2(a). Here the motion of electron creates field deficit to ions at the rate of its cyclotron frequency ($\Omega_e$). Due to these reasons, ion damping is a high frequency damping but electron is a low energy one. When the ions and electrons are coupled in \emph{\textbf{lhos}},  at the initial stage of coupling, the Larmour radius of electron ($r_e$) is smaller than ion ($r_i$); now the particles are at far field positions and the electrons spiral inside $r_i$. In each rotation of ions, it gains energy by electron damping and  $r_e$ increases. Here one point is to be noted; the energy absorbed must be equal to field deficit; when $r_e < < r_i$, the particles are at far field positions, so field deficit is small and the rate of absorption also small. Thus the electron damping is a gradual and low energy one and this continues till $r_e  \approx r_i$, here the ion Landau damping starts. These points can be confirmed from the graph of \emph{\textbf{gire}} shown in the fig.3 \citep{Antony2017a}. The first part of plot shows a positive slope, this is due to the damping of accoustic waves, here \emph{\textbf{emws}} gain energy. The second part, the slope is negative, now the damping  of \emph{\textbf{emws}} starts. In this region, the damping is gradually increasing with frequency, starting from a low energy waves. This is exactly reflecting the properties of electron damping at the initial stage of coupling in the \emph{\textbf{lhos}}, just as the particles are at far field positions, the field deficit and energy absorptions are small. Therefore  we can confirm the initial damping is due to the photon absorption by electrons, since the frequency is unique at every point in the graph and increases with energy, therefore this is a single photon absorption process.

\begin{figure}
	\centering
	\includegraphics[height=7cm]{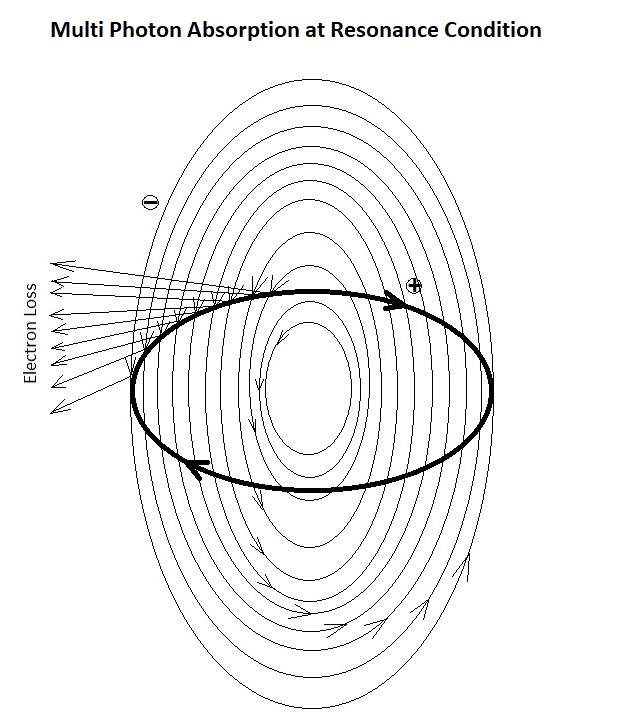}
	\caption{Figure shows the multi photon absorption at resonance condition due to the Deficit of Electrons. Since $\Omega_e >> \Omega_i$, electrons meet ion at every point of its orbit and this gives continuous field deficit. The resonance condition, $r_i = r_e$ is satisfied at every point in the postphase region where ion Landau damping is possible.  The resonant wavelength of photon is proportional to the Larmour radius, since the electrons are creating the field deficit  continuousely in the same orbital length, the ions absorb several photons of same energy.}
\end{figure}

\subsection{Ion Landau Damping}
Ions are inertial, a high field is required to accelerate them and this happens at resonance condition, when $r_e = r_i$, the particles are much closer and experiences intense electric field. At this condition, a small variation in the field causes heavy acceleration to both the particles. The electrons are already in an accelerated mode by prephase damping and creates a field deficit at the rate of $\Omega_e$. An important point is that these elelctrons are highly relativistic and at the edge of $r_i$, they are easily lost out from the potential of ions. This makes the field deficit more intense which gives multiple acceleration to ions in the direction of photons. Thus the rate of damping extremely high at resonance condition, which can be seen from the graph (see fig.3). One of the most important point is this; since $\Omega_e >> \Omega_i$, electrons meet the ions at every point of its orbit and gives continuous field deficit in the same orbital length. This situation is depicted in fig.4, here the resonance condition, $r_i = r_e$ is satisfied at every point of the same orbit, therefore the ions continuousely absorb photons of same amplitude (energy). We find the deficit is starting from the extreme end of the ion orbit where highly relativistic electrons are scattered out and then it comes to lower energy electrons too. So this gives a continuity in the resonance and most of the electrons are lost due to this high field situation creates the field deficit quite critical to ions which under go multiple acceleration at the same orbital length (same wavelength) by absorbing sufficient number of photons. This is can be confirmed from the graph of \emph{\textbf{gire}} (see fig.3), in the last pasrt of damping where frequency is not changing with damping and this is similar to multi photon absorption.  Here the loss of electrons at resonance is compensated by photo-electrons emitted during \emph{\textbf{mpi}}. \\
one of the important outcome of prephase Landau damping is, loss in the electron density. Since they are lighter particles easily get accelerated and escape from the ion potential. Due to this reason,  the electrons cannot be accelerated to higher energies and maximum possible acceleration must be less than the free electron energy (511 keV). This is also only at bound state, i.e., in \emph{\textbf{lhos}} when $r_e \leq r_i$. This is one of the most crucial information regarding acceleration of ions in solar corona. The deficit of electrons is a continuous phenomenon in \emph{\textbf{lhos}} and it is further amplified by coronal loop oscillations \citep{Antony2017b}. 

For giving a clarity, we list the differences between electron and ion Landau Damping in table form (Table 1)

\begin{figure}
	\centering
	\includegraphics[height=7cm]{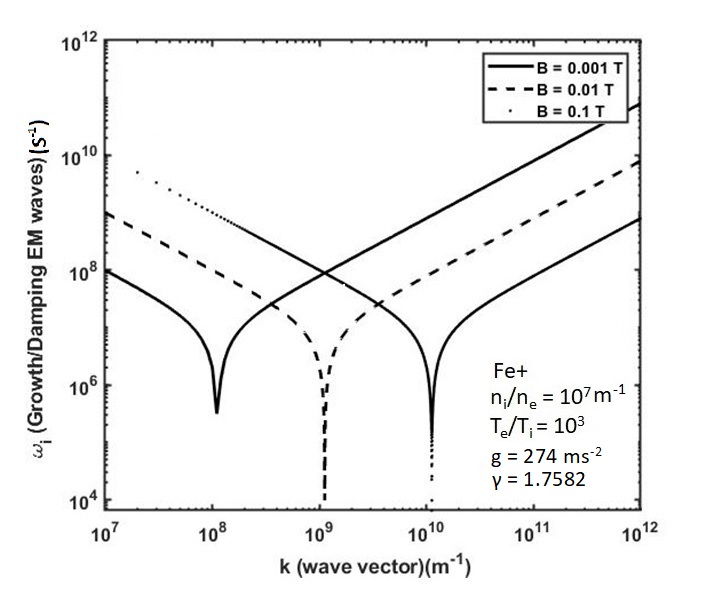}
	\caption{\textbf{Effect of $\mathbf{B}$ over  \emph{\textbf{mpi}}}: Logarithemic plot of damping/emission \emph{\textbf{emws}} with wave vector at resonance condition. It shows \emph{\textbf{mpi}} increasee with $\mathbf{B}$ both in magnitude (number of photons) and frequency}.
\end{figure}

\subsection{High Field at Resonance} We understands from the above discussion, $3^{rd}$ and $4^{th}$ conditions for \emph{\textbf{mpi}} are satisfied in \emph{\textbf{gire}} at resonance condition. But for a comparative analysis we shall do an average calculation on the rate of energy  absorption in \emph{\textbf{gire}}. Let ${\nu}_r$ be the resonant frequency, then the energy absorbed by an ion in this frquency is $h{\nu}_r$.  This energy is absorbed at time of $t =  \frac{1}{{\nu}_r}$ and since the frequency is not changing during resonance condition, then energy absorbed per second  is given $Es^{-1} = h{{\nu}_r}^2$,
 Inserting the value of resonant frequency\cite{Antony2017b},

$$Es^{-1} = (\frac{h c^2 e^2 }{4{\pi}^2 m M}) (\frac{Z}{\gamma W}) (\frac{n_i}{n_e} \frac{B^2}{gSin\theta}) $$

Where \textbf{h}, Planck's constant; \textbf{e}, value of electric charge; \textbf{c}, velocity of light; \textbf{m} \& \textbf{M} are masses of electron and proton; \textbf{Z}, degree of ionization, \textbf{W}, mass number, $\gamma$, Lorentz factor, $n_i$  is the number density ions drifted per meter; $n_e$ is the number density of electrons; $\mathbf{B}$, magnetic field; $\mathbf{g}$, acceleration due to gravity.
There are several varibles in this formula, since we are comparing it with the maximum power of Laser, we shall choose conditions of corona, where the hard X-rays are observed.  Keeping, a high $\mathbf{B}$, for lower corona where the coupling starts (we take $\mathbf{B} = 0.1$T, but $\mathbf{B}$ can go upto 0.25T); and field deficit   $\frac{n_i}{n_e} = 10^7$ is chosen by assuming all the coupled electrons are lost out at resonance condition, also considering the elemental abundance. We keep the angle $30^{o}$ between the $\mathbf{B}$ and g for lower corona. Using these values subtituting all constants we can calculate the power as $Es^{-1} = 1.16359 \times 10^{23} (\frac{Z}{\gamma W})$ eV$s^{-1} $. This value is very high and sufficient for ionizing any atom, but the rate of absorption must be analyzed in the atomic time scale. Ionization is a quantum mechanical process and the energy absorbed is unique for every element, this must be analyzed on the basis of their ionization time. 

\begin{figure}
	\centering
	\includegraphics[height=7cm]{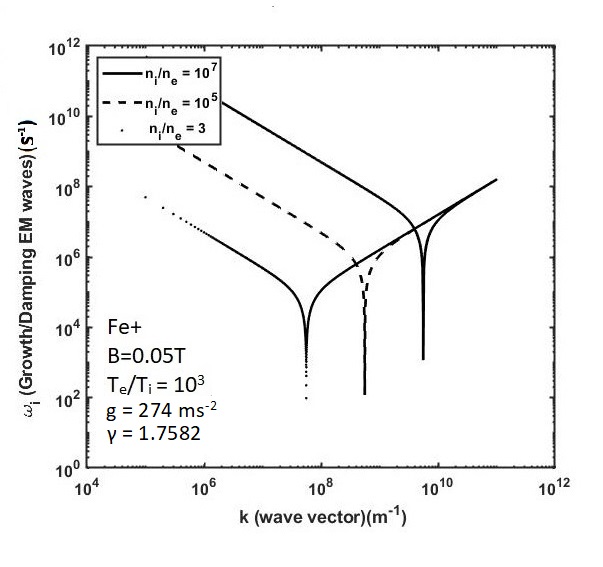}
	\caption{\textbf{Effect of Electron deficit in \emph{mpi}}: Figure shows the effect of field deficit in resonant condition. The rate of absorption increases in frequency with field deficit }.
\end{figure}

\subsection{Temporal Compression by Electron deficit}  The fourth condition is satisfied by the  deficit of accelerated electrons at resonant conditions. In laser ionization, time compression is obtained by compressing photons into high intensity pulse in short duration. This is done by grating device, which gives definite phase shift to various beams of same frequency and made to converge at single point.  The same effect is produced by a bunch of highly relativistic electrons in the lower hybrid coupling. Since $\Omega_e >> \Omega_i$, electrons meets ion at every point of its orbit gives a continuos acceleartion to ions which are in phase with photons of same frequency. Fig.4 shows the situation, since electrons are lighter particles, easily lost out and this loss creates the field deficit more critical to ions, starting from the edge of ion orbit, the electrons continuousely creating this situation at every point in the same orbit of ions. \textbf{The resonant wavelength of photon is proportional to Larmour radius, since the electrons creating field deficit continuousely in the same orbital length of ions, they absorbs the same energy photons.} The degeneracy of electrons also may be helpfull in this regard for getting a continuous acceleration at resonance condition. Each electron is unique with its energy and orbits in \emph{\textbf{lhos}} and the difference gives a continuity in field deficit at resonance.

\begin{figure}
	\centering
	\includegraphics[height=7cm]{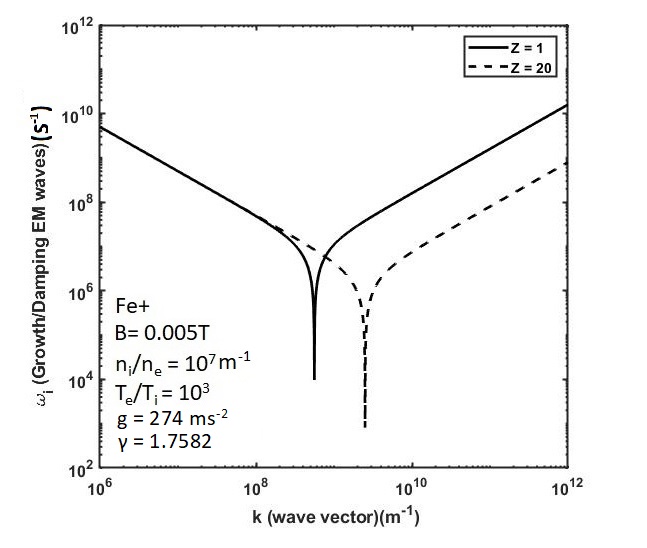}
	\caption{\textbf{Effect of ionization (Z) over Ion damping}: Figure shows the variation in Damping rate over degree of ionization. We find multi photon absorption increases with degree of ionization}.
\end{figure}
  
 \section{Effect of various parameters over \emph{mpi} in  \emph{gire} }
 
 At resonance, the ions are undergoing \emph{\textbf{mpi}} and possibly emits all types of radiations.  Unlike in laser, \emph{\textbf{gire}} has a number parameters are playing very important role in the resonance condition and we briefly discuss in this section.  
   
   \subsection{Effect of B, $\frac{n_i}{n_e}$ \& g on Resonance}
     \textbf{g}: Among these 4 variables, g is a static variable depends on the direction of $\mathbf{B}$; the effective value depend on where the coupling occurs. Generally this happens at lower corona or coronal loop where $\theta$ is small  $< 30^{\circ}$. For upper corona  $\theta \approx 90^{\circ}$ ion damping will be less comparatively at lower part of the loop. The value of \textbf{g} appears in the denominator, this reduces the rate of damping and this may be the reason that why most of hard X-rays are observed in the lower corona where the effective value is less. \\ 
   
  \textbf{B}: This is one of the most sensitive variable which affects the resonance conditions more than any other. If we look into the formula,  $\mathbf{B}$ appears in square times magnitude compared with other variables, is controlling the resonance condition. This is quite meaningful to coronal conditions, since all the dynamical structures of corona is controlled by $\mathbf{B}$. The structural variations in the $\mathbf{B}$ has immense effect on heating corona, which has been observed for the last 75 years. At low $\beta$, particles are in cyclotron motion and the rate of resonant damping depend on rotation frequencies.  Since they are rotating in opposite direction, even small variation in $\mathbf{B}$, increases the frequencies for both the particles in a logarithemic scale which results in the high damping. This makes the heating more easier in high $\mathbf{B}$ regions and this result is established in the fig.5., whichfigure shows, both magnitude and frequency increases with $\mathbf{B}$, here the plot is done with the expression of growth/damping for \emph{\textbf{emws}} \citep{Antony2017a} \\
    
  \begin{figure}
  	\centering
  	\includegraphics[height=7cm]{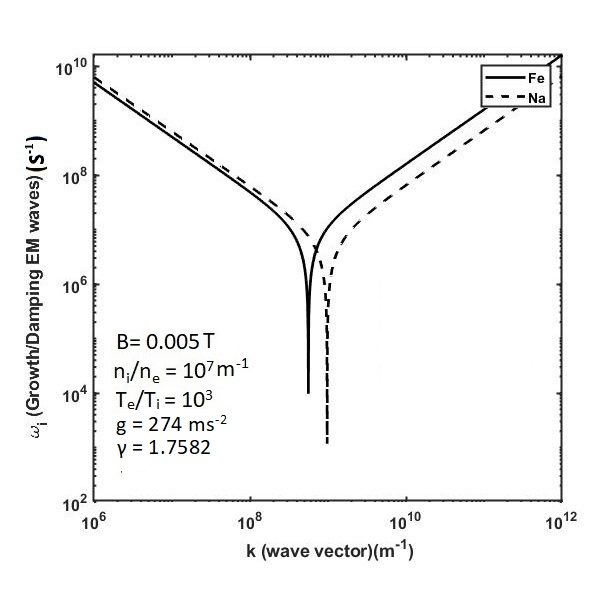}
  	\caption{\textbf{Effect of mass number over \emph{mpi}}: Resonance plot reflects the effect of mass number; shows the damping decreases with mass.  For heavier elements, the resonance frequency and damping both are decreasing}.
  \end{figure}
   
  \textbf{$\frac{\mathbf{n_i}}{\mathbf{n_e}}$}: We have eloborately discussed the role of field deficit which makes the multi photon absorption possible. The density of electron is inversely related to $\mathbf{B}$. In high \textbf{B}, the $r_e$ is small, therefore the loss of electrons are less which makes the electrons acceleration is possible to maximum energy. This suggests that at high \textbf{B}, the electron scattering at resonance will be quite high in strong \textbf{B} region, this supports the observations in coronal emissions of hard X-rays. 
  For upper corona, the magnetic field is less where the electrons loss is maximum which gives the deficit is high. Here electrons are lost due to prephase damping and continues till the resonance condition where it reaches maximum and this result has been estabilished in the fig.6. 
 We find the damping date increases in both frequency and magnitude. 
 So combining these two results we can conclude this, when the \textbf{B} is high the {$\frac{\mathbf{n_i}}{\mathbf{n_e}}$} is less, and damping rate controlled by \textbf{B} or if \textbf{B} is low, {$\frac{\mathbf{n_i}}{\mathbf{n_e}}$} is high and the damping is high. Thus both are reciprocally controlling the damping rate.
       
    \begin{figure}
 	\centering
 	 	\includegraphics[height=6.5cm]{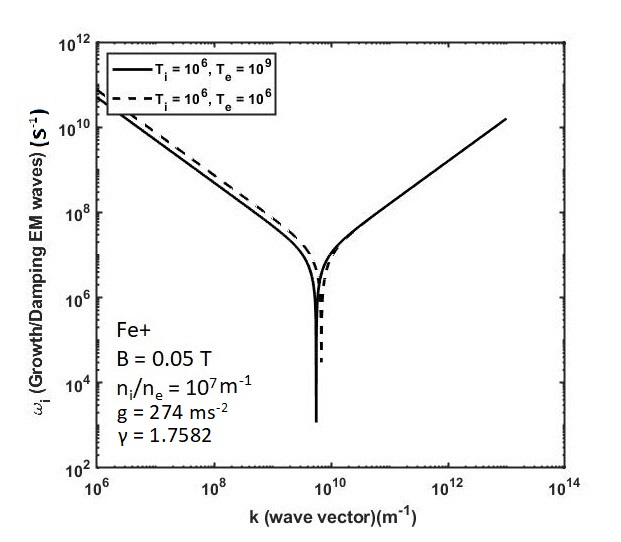}
  	\caption{ \textbf{Effect of Temperature over resonant damping}: Figure shows the rate of damping at resonance condition for difference in Temperatures of ion ($T_i$) and electron($T_e$). Resonance is maximum when the particles are closer and this happens when the Larmour radii are same. For this, the temperatures must differ by an order of their mass ratios, i.e., $ \approx10^3$ which is established in this graph.}.
 \end{figure}

\subsection{Effect of Z \& W}
The charge state of ion is continuousely changing during  \emph{\textbf{mpi}} and rate of damping increases with Z due to higher couloumb force . This result is established in the fig.7. We find resonant damping enormously increases with Z and this is helpful for getting continuity in ionization process. Regarding mass number, we find damping is reduced for heavier elements but one of the important benefit is the frequency shift. For higher elements, the resonance condition shifts to low energy region where the maximum intensity is possible due high solar luminosity. For lighter elements damping is quite high and due to this reason, these elements will be easily disintegrate. This is reflected in the fig.8., which shows the ion damping for Fe+ low compared with Na+.

\subsection{Effect of Temperature}
One of the most important parameter that we are interested is temperature but it is invisible and reflected in the relativistic factor $\gamma$. The particles resonance is possible only when the larmour radii are same; imposing this condition we can deduce a relation between electron and ion temperatures \citep{Antony2017b};
$$T_i = \frac{Z^2}{W }\frac{\gamma m}{M}T_e$$
where $T_i $ and $T_e$  temperatures,  rest al the parameters were mentionede earlier.  $\frac{T_e}{T_i } \approx \frac{M}{m}$, i.e., $T_e$ must be greater atleast by factor of $10^3$ to get maximum resonance and this is established in the fig.9. 
  
 \section{Conclusion} 
   A comparative study of \emph{\textbf{mpi}} in Laser and \emph{\textbf{gire}} is done. Both are same except prephase electron damping before the resonance condition. In \emph{\textbf{gire}}, the whole process is archestrated by a bunch of electrons accelerated by Landau damping. The temporal compression of photons is obtained by grating element but in \emph{\textbf{gire}} it is created by field deficit due to the loss of electrons at resonant condition. This loss is compensated by the photo electrons emmitted by  \emph{\textbf{mpi}}.  In Laser ionization, absorption/emission specrtum is very narrow, but in \emph{\textbf{gire}}, it is wide and continuous range of all frequencies. The laser ionization is solely depend on the intensity (number of photons) but \emph{\textbf{gire}} has seven variables with more tunability and here the process is controlled by $\mathbf{B}$ with electrons, hence the high spectral resolution will be high.
 
\textbf{Acknowledgment}
This work is done under UGC Research Award Scheme (2016-2018) and acknowledges the help of Sunitha Ambrose in programming part.

\end{article} 
\end{document}